# Anomalous thickness-dependent electrical conductivity in van der Waals layered transition metal halide, Nb$_3$Cl$_8$

Jiho Yoon[*1], Edouard Lesne[1], Kornelia Sklarek[1], John Sheckelton[2], Chris Pasco[2], Stuart S. P. Parkin[1], Tyrel M. McQueen[2], and Mazhar N. Ali[*1]

[1]*Max Planck Institute of Microstructure Physics, Weinberg 2, Halle 06120, Germany*
[2]*Department of Chemistry, The Johns Hopkins University, Baltimore, MD 21218, USA*

## Abstract

Understanding the electronic transport properties of layered, van der Waals transition metal halides (TMHs) and chalcogenides is a highly active research topic today. Of particular interest is the evolution of those properties with changing thickness as the 2D limit is approached. Here, we present the electrical conductivity of exfoliated single crystals of the TMH, cluster magnet, Nb$_3$Cl$_8$, over a wide range of thicknesses both with and without hexagonal boron nitride (hBN) encapsulation. The conductivity is found to increase by more than three orders of magnitude when the thickness is decreased from 280 μm to 5 nm, at 300 K. At low temperatures and below ~50 nm, the conductance becomes thickness independent, implying surface conduction is dominating. Temperature dependent conductivity measurements indicate Nb$_3$Cl$_8$ is an insulator, however the effective activation energy decreases from a bulk value of 310 meV to 140 meV by 5nm. X-ray photoelectron spectroscopy (XPS) shows mild surface oxidation in devices without hBN capping, however, no significant difference in transport is observed when compared to the capped devices, implying the thickness dependent transport behavior is intrinsic to the material. A conduction mechanism comprised of a higher conductivity surface channel in parallel with a lower conductivity interlayer channel is discussed.

# Introduction

The discovery of graphene by exfoliation allowed a plethora of new physics, like the quantum Hall effect [1,2], and massless Dirac fermions [3], to be explored in van der Waals (vdW) layered two-dimensional (2D) systems. Through both simple mechanical exfoliation, and the dry transfer method via adhesive tape [4, 5], various 2D vdW layered materials (including transition metal dichalcogenides (TMDC), and transition metal halides (TMH)) have been reinvestigated not only bulk, but also in the 2D limit (approaching monolayer). Thickness dependent changes to properties such as topological state and the presence of massless Dirac, and Weyl fermions [6, 7, 8], onset of charge density waves (CDW) [9], as well as superconductivity [10, 11], have all recently been observed. Furthermore, heterostructures of TMDCs/TMHs where consecutive monolayers are stacked at an angle with respect to the underlying layer, or interfaced with other 2D materials have also shown unexpected emergent properties including a correlated insulating state [12], interfacial superconductivity [13], and anomalous Hall effects [14]. In particular, 2D TMHs, which consist of vdW layers made up of transition metals like Nb, Cr, Mo, Ru, etc., and halides (I, Cl, Br), have received much attention due to their promising potential in a wide variety of magnetic phenomena based on ferromagnetic [15], and antiferromagnetic insulators [16, 17], as well as geometric frustration linked to prospective quantum spin-liquid behavior [18, 19, 20]. Therefore, efforts to reveal new types of 2D TMHs with potentially exotic magnetic properties, have rapidly increased in the last few years.

$Nb_3Cl_8$, in particular, is a 2D vdW layered TMH cluster magnet insulator, which has been studied primarily with regard to its structure; and contains Nb-Nb triangular clusters resulting in an insulating ground state of localized electrons [21, 22, 23]. The crystal structure of $Nb_3Cl_8$ is strongly anisotropic; with planes of Nb atoms octahedrally coordinated by Cl, stacked upon each other along the c-axis as shown in figure 1(a). In plane, the Nb atoms make up a triangular lattice of Nb-trimers which form magnetic clusters with a net effective spin $S_{eff}$ =1/2 in each $Nb_3$ unit. There are two $Nb_3Cl_{13}$ clusters per unit cell, and the stacking order of layers is alternated as -AB-AB- sequences, due to the inequivalent interfacial-capped Cl atoms, which differs from conventional TMDCs [24]. The surface stability, structures, and defect states have been further examined through atomic force microscopy (AFM), and scanning tunneling microscopy (STM) to understand defect states [25].

Recently, $Nb_3Cl_8$ has regained the attention of both experimentalists, and theoreticians owing to its potentially geometrically frustrated magnetic structure. The so-called pseudo Kagome lattice formed by the Nb-trimer cluster in the ab-plane of the structure has a single localized unpaired spin which couples antiferromagnetically with its neighbor. $Nb_3Cl_8$ has been experimentally reported to show a magnetic phase transition between a high temperature paramagnetic, and nonmagnetic singlet state at 90 K, arising from the layer rearranging structural phase transition, which lowers the symmetry from trigonal $P\overline{3}m1$ to monoclinic C2/m [26, 27]. Two possible explanations of the singlet state are based on either a second-order Jahn Teller (SOJT) distortion [26] or charge disproportionation in adjacent layers [27], but a consensus has not been reached. $Nb_3Cl_8$ has also been predicted to be a potential monolayer ferromagnet [28], and topological insulator [29], although monolayer kagome nets also have the potential to host a spin liquids ground state, alike $\alpha$-$RuO_3$ [35].

While there have been several experimental, and theoretical studies on $Nb_3Cl_8$, its electrical transport properties, particularly as a function of thickness, have not yet been reported [33]. Here, we report the temperature, and thickness dependent electrical conductivities of high-quality single crystal flakes of $Nb_3Cl_8$ (within the measurable range of the experiment) ranging from 280 μm to 5 nm. We find an unexpected trend of increasing conductivity, and decreasing activation energy as the thickness is decreased, while below ~50 nm, the conductance is found to be thickness independent. These results strongly imply the presence of enhanced surface conduction with an insulating bulk, and is discussed in terms of a parallel channel conduction model [31, 33].

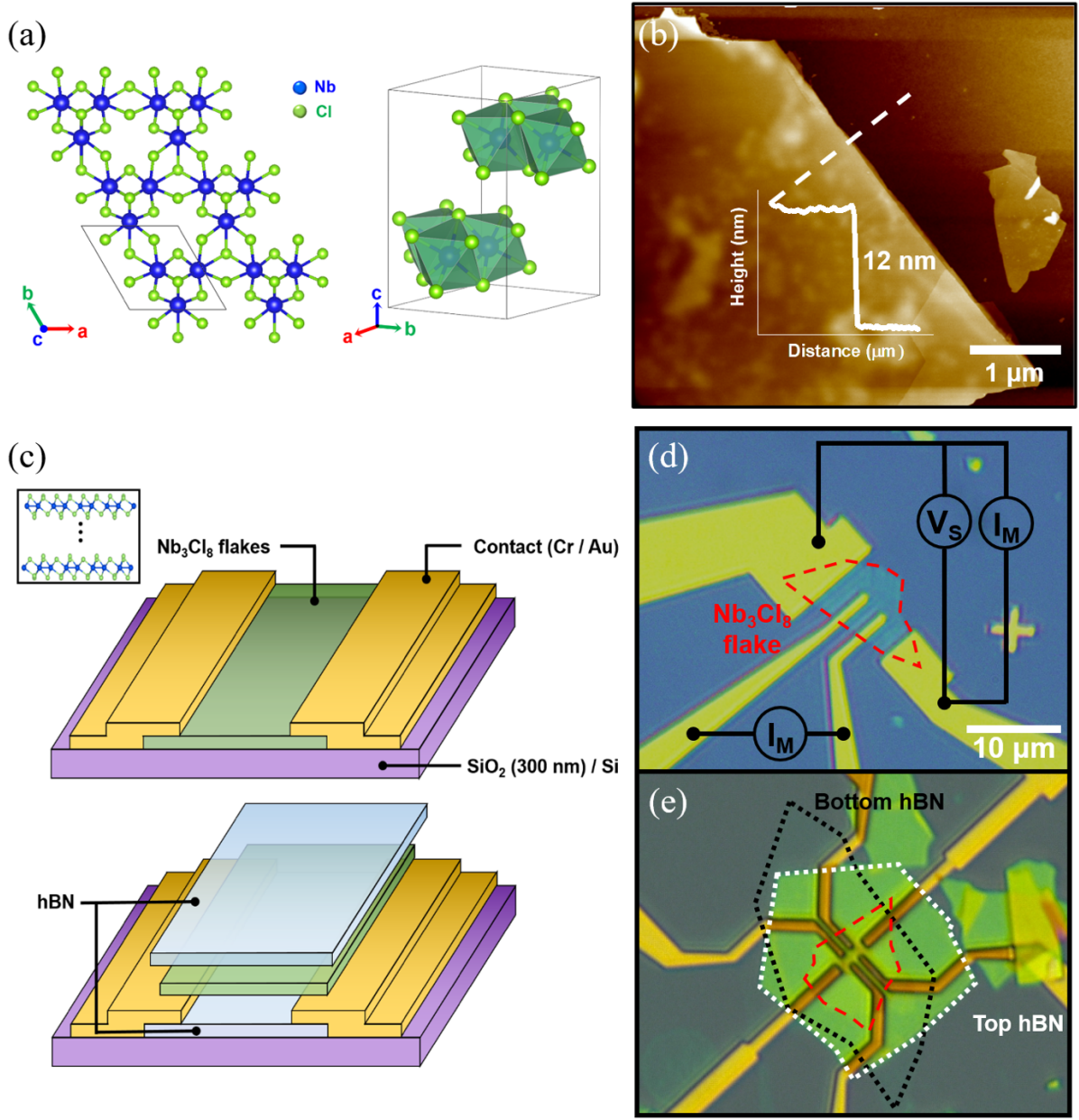

FIG. 1. (a) Crystal structure of $Nb_3Cl_8$, showing transition metal (Nb) trimers forming a triangular lattice of $S_{eff} = 1/2$ in the ab-plane. The $Nb_3Cl_8$ layers form an -AB-AB- stacking sequence. (b) AFM image of a mechanically exfoliated 12 nm thick flake on the $SiO_2/Si$ substrate. (c) Schematics of non-hBN encapsulated device (top), and hBN encapsulated device (bottom). (d)(e) Optical microscope images of an uncapped and capped devices, respectively. The red dashed line shows the outline of $Nb_3Cl_8$ flake. The black (white) dotted line encloses the bottom (top) layer of hBN, respectively. Circuit elements show the two-point and four-point voltage source ($V_S$), and current measurement ($I_M$) configurations.

# Experimental (Method)

High-quality single crystals of $Nb_3Cl_8$ were used for multilayer flake preparation. Flakes of $Nb_3Cl_8$ with varying thicknesses were prepared by conventional mechanical exfoliation with adhesive tape, and carried out in a dry nitrogen glove box. The thickness of the $Nb_3Cl_8$ flakes was determined by optical microscopy, and atomic force microscopy (AFM) as shown in figure 1(b). For the 280 µm thick sample preparation, four in-line contacts were manually made directly on the bulk crystal with silver paint, and further attachment of gold wires. Thinner flake devices were prepared as follows: exfoliated flakes were transferred onto insulating $SiO_2/Si$ substrates (300 nm thick $SiO_2$ layer). Electron beam (e-beam) lithography and lift-off processes with thermal deposition of Cr/Au were used for four-point contact fabrication. In order to prepare devices with hBN encapsulation, hBN layers were first transferred onto the $SiO_2/Si$ substrates, then contacts were patterned onto that hBN bottom layer as shown schematically in figure 1 (c). Suitable $Nb_3Cl_8$ flakes were identified by optical microscopy, and then a propylene carbonate (PC) stamp with the poly-dimethyl siloxane (PDMS) transfer method was applied for creating the sandwich structures with a second hBN flake capping the exfoliated $Nb_3Cl_8$ layer [4]. Thus, we have successfully prepared hBN encapsulated $Nb_3Cl_8$ devices in inert conditions, without the need for further chemical processing. Typically, 10-20 nm thick hBN flakes were used as encapsulation layers. Figures 1(d), and 1(e) show optical images of non-encapsulated, and encapsulated devices, respectively. The electrical conductivity was measured with four-, and two-point probe methods in a Quantum Design PPMS combined with a Keithley 6430 sub-Femtoamp Remote Source Meter (equipped with constant voltage source, and current measurement units). The temperature range of 300 K through 180 K was investigated due to detection limitations of the electrical detection setup (when the resistance of devices is found to exceed hundreds of GΩ below 180 K) equipment. XPS spectra were obtained from pristine, and e-beam resist processed flakes in a Thermo Scientific K-alpha XPS system with a monochromatic Al Kα source. The in-situ cluster mode Ar ion etching was used for investigation of surface contamination.

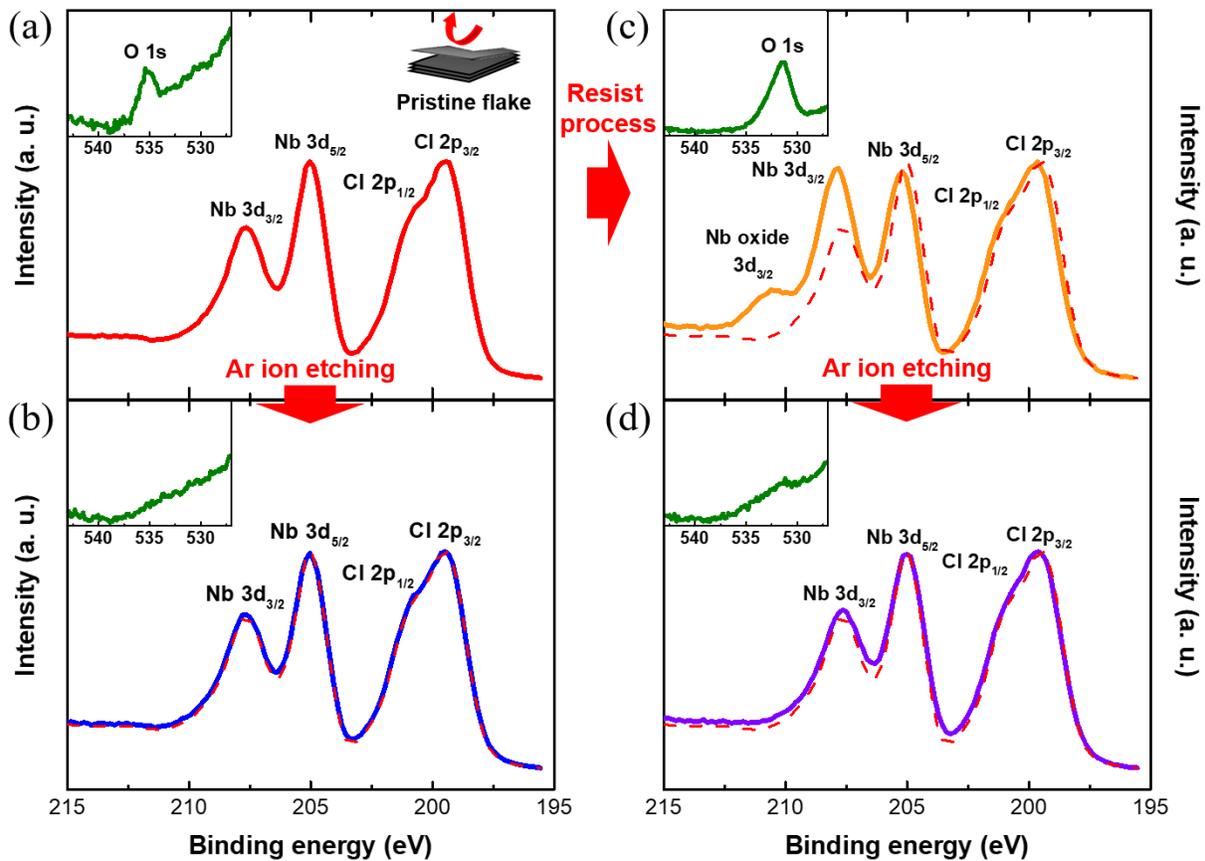

FIG. 2. XPS spectra of (a) pristine $Nb_3Cl_8$ flake, freshly exfoliated by adhesive tape. Nb 3d, and Cl 2p core level peaks (red solid) are visible. The inset displays oxygen 1s peak (green solid). (b) Nb 3d, and Cl 2p core level peaks in the spectra (blue solid line) after a low-energy Ar ion etching process with the original pristine flake spectrum overlayed (red dashed line). Inset shows no oxygen peak. (c) XPS spectrum after the standard resist processing (orange line). Niobium oxide features, and a shifted oxygen peaks (inset) are observed. (d) Spectrum taken after an identical Ar ion etching process as (b), showing the absence of the Nb oxide peak in the main panel, and a very faint oxygen peak in the inset.

# Results and discussion

In order to confirm that the chemistry (and correspondingly the properties) of the samples were not significantly affected by the lithography process, the oxidation, and potential contamination from the lithography process on the surface of a fresh surface of $Nb_3Cl_8$ were investigated via XPS measurements. Since performing XPS measurements on the fabricated devices was not possible due to their small form factor, and lateral resolution of the setup (limited by the incoming beam spot size), XPS was performed on the surface of the single crystal, before, and after it was exposed to the photolithography process, as an analog to the $Nb_3Cl_8$ fabricated devices. Figure 2 (a) shows the Nb 3d, and Cl 2p core spectra of pristine bulk crystal, after it was freshly cleaved by adhesive tape. The Nb 3d spectra shows two peaks for the $3d_{5/2}$, and $3d_{3/2}$ states at 205 eV, and 208 eV, respectively, due to spin-orbit coupling splitting. However, in the Cl 2p spectra, the $2p_{3/2}$, and $2p_{1/2}$ peaks at 199 eV and 201 eV, respectively, are partially convoluted, and not as clearly identifiable. In addition, a faint feature corresponding to oxygen 1s peak is visible at 535 eV (inset of Fig. 2(a)). Note that no shift of the Nb 3d, and Cl 2p peaks, compared to the previous report, is evident [23]. Ar cluster ion etching was employed to remove surface contamination, and confirm that the oxygen impurity was not present deeper into the $Nb_3Cl_8$ crystal; as shown in figure 2 (b). The total etched thickness of $Nb_3Cl_8$ was estimated to be less than 1 nm, and while it is clear that the etching process does not disturb the Nb 3d, and Cl 2p peaks, the O 1s peak is however no longer visible (see inset), which demonstrates that $Nb_3Cl_8$ is stable against oxygen in atmosphere, and that any oxidized region is limited to a thin self-passivating layer as its surface.

After the e-beam lithography process, however, the appearance of an $NbO_x$ 3d peak at ~210.5 eV, as well as a shift of the Nb 3d, and Cl 2p peaks were observed as shown in figure 2(c).

Unlike in figure 2(a), the O 1s peak appears at around 532 eV, which indicates that the oxygen bonded as a metal oxide or metal carbonate, and the Nb 3d and Cl 2p peaks are shifted about 0.5 eV higher in binding energy compared to the original pristine sample. To investigate the surface oxidation depth, Ar cluster ion etching was carried out in the same conditions as previously described. Figure 2(d) shows the spectra after etching, and the Nb 3d and Cl 2p peaks are restored to the original state along with the disappearance of the NbO$_x$ 3d spectra. Therefore, we can conclude that oxygen does not significantly penetrate inside of the crystal or flake as a result of our patterning protocol.

The electrical conductance (G), and conductivity (σ) of Nb$_3$Cl$_8$ flakes of different thicknesses, ranging from 280 μm to 5 nm thick, was measured in a constant voltage source mode in a four-probe configuration as shown in figure 1(d). Figure 3(a) displays the electrical conductance as a function of thickness ($t$) in a semi-log scale at various discrete temperatures from 300 K to 180 K. As expected for an insulator, the conductance universally decreases as temperature decreases. However, as $t$ of the samples is varied, a more complex, and intriguing behavior emerges. At all temperatures, G decreases as $t$ is reduced from bulk to several hundreds of nanometers. However, by $t$ ~50 nm, G exhibits a jump compared with the ~330 nm data point. It is unclear what the origin of this jump in conductance is, although it may be related to the onset of finite size effects. As we further reduce the thickness, we see two different types of responses of G depending on the temperature. Above ~200 K, the conductance once again decreases monotonically with $t$, but below 200 K, G appears to be - thickness independent -. Decreasing $t$ no longer decreases the amount of current carried by the flake. This implies that below $t$ ~50 nm, and 200 K, the Nb$_3$Cl$_8$ reaches an apparent limit where its "surface" (or a thin sheet) is carrying most of the current in place of the more insulating bulk of the flake.

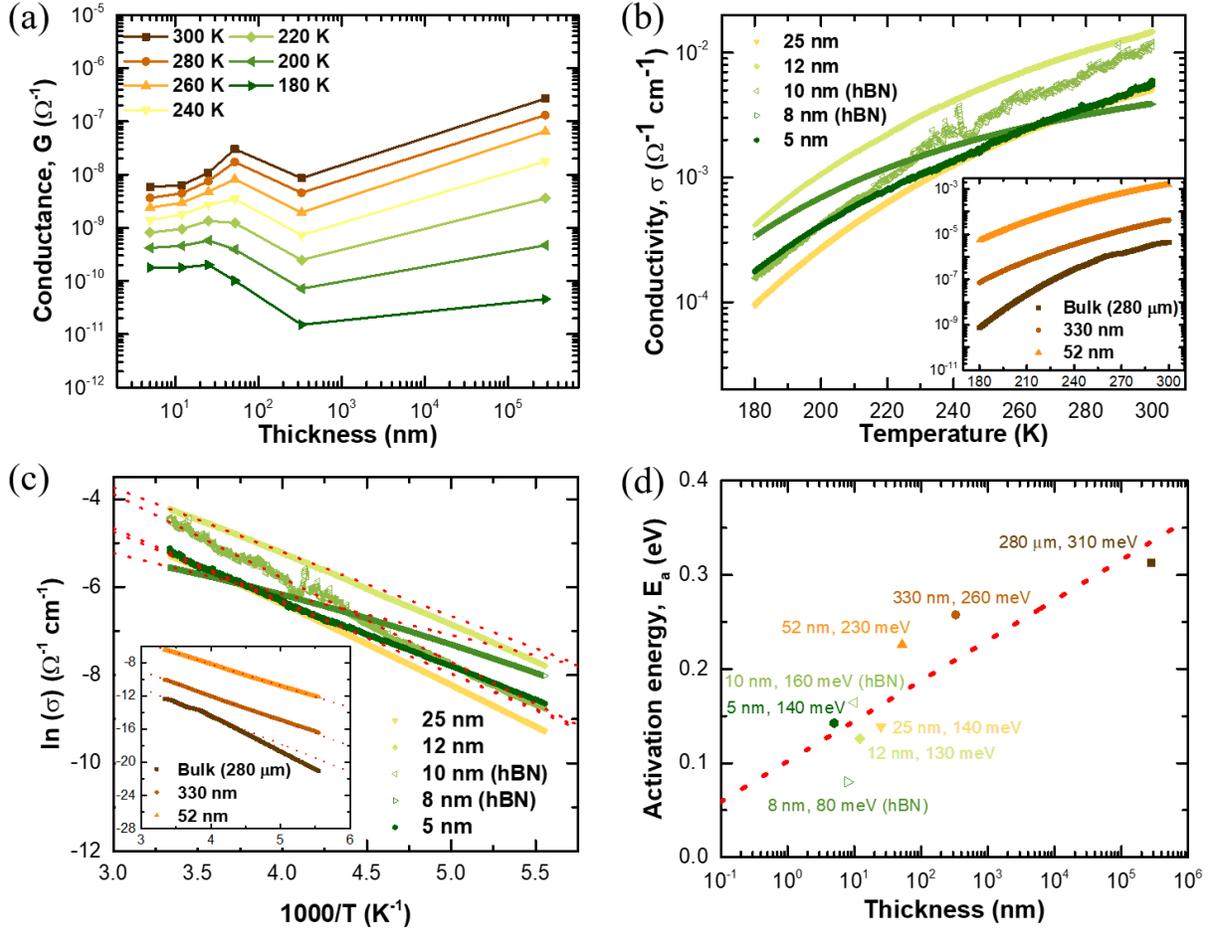

FIG. 3. (a) The electrical conductance of $Nb_3Cl_8$ flakes ranging from 5 nm to 280 μm thick. The conductance is compared at various discrete temperatures from 300 K (brown square) to 180 K (dark green triangle), every 20 K steps. (b) The log plot of conductivity as a function of temperature from 300 K to 180 K for different flakes thicknesses. hBN capped flakes of 10 nm (light green open triangle), and 8 nm (green open triangle) are also displayed. (c) Arrhenius plot of the conductivity of various $Nb_3Cl_8$ flakes. The red dashed lines correspond to the best fit of the high temperature region, according to the activation energy equation (see main text). (d) The activation energy $E_a$ as a function of flake thickness is shown in semi-log plot, with best fit shown by the red dash line.

Figure 3(b) displays the electrical conductivity as a function of temperature for different flakes thicknesses. The temperature dependent conductivity verifies $Nb_3Cl_8$ as an insulator under the measureable investigated range, regardless of thickness. In principle, the conductivity of most materials is constant with respect to their dimension until the limit of finite size imposes boundary conditions on the periodic potential generated by the crystal structure; in which latter case the conductivity is generally found to decrease as a result of increased surface, and boundary induced scattering [32]. For $Nb_3Cl_8$, however, we find an anomalous thickness-dependent conductivity. Indeed, the semi-log plot clearly reveals that the conductivity increases more than three orders of magnitude from $10^{-6}$ $\Omega^{-1}cm^{-1}$ to $10^{-3}$ $\Omega^{-1}cm^{-1}$ at room temperature (300 K) upon decreasing of the $Nb_3Cl_8$ flake thickness from 280 μm to 5 nm. Previously, there have been reports of changes in conductivity with thickness variations in three-dimensionally grown thin films by induced strain [33], and in very few cases of two-dimensional layered system, such as TMDCs, like $MoS_2$ [30], and graphene multilayers [31], however such a phenomenon had not yet been reported in TMHs.

Through the comparison of the conductance and the conductivity results, we find that the thickness-dependent conductivity is, remarkably, an intrinsic property of $Nb_3Cl_8$. The conductance G is expected to linearly depend on the conductivity σ according to the relation: $G = \sigma \frac{wt}{l}$, where $w$, $l$, and $t$ are the width, length, and thickness of the conducting channel, respectively. By comparing figures 3(a), and 3(b), we observe that change in conductivity are significant despite relatively small changes in conductance. As the variations of the dimension factor $w/l$ in lithographed devices are kept distinctively small (within a factor of 2), the observed large changes in conductivity implies that the thickness $t$ is the major factor in determining the change in conductivity of $Nb_3Cl_8$. To further rule out the effect of the surface oxidation, and impurities, hBN encapsulated $Nb_3Cl_8$ flakes with 8 nm, and 10 nm thick were simultaneously investigated and show the same pattern as the non-encapsulated thicker flakes. This observation corroborates that the anomalous change of conductivity in thickness is an intrinsic property of $Nb_3Cl_8$ thin flakes, and cannot be attributed to the device fabrication procedure.

From the temperature-dependent conductivity measurements, the termal activation energy $E_a$ of the carriers, corresponding to the energetic band gap, was obtained by fitting the Arrhenius equation $\sigma(T) = \sigma_0 \exp\left(\frac{-E_a}{kT}\right)$, where T is temperaure, $\sigma_0$ is the residual conductivity, and $k$ is the Boltzmann constant. Figure 3(c) shows the Arrhenius plot of the natural logarithm of conductivity as a function of 1000/T, and $E_a$ is obtained from the slope of the high temperature region. As a result, the semi-log plot of the thickness-dependent $E_a$ is shown in figure 3(d). According to the previous report, the activation energy of $Nb_3Cl_8$ pellets was found to be ~210 meV via eletrical conductivity measurements [30], which is smaller  than the 310 meV we extracted for the bulk single crystal ($t = 280$ μm) in our study. This may be due to the pellet sample having greater defect concentration or slight doping which increased its conductivity. We also observe that $E_a$ decreases as the thickness of $Nb_3Cl_8$ flakes is decreased by about one half, from 310 meV  to 140 meV ($t = 5$ nm), and is about linearly correlated to log $t$.  Based on this linear fitting, the $E_a$ of a monolayer of $Nb_3Cl_8$ ($t = 3.446$ Å) is estimated to be approximately 50 meV; which greatly differs from the theoretical predicted value of 638 meV [28]. This large quantitative difference between theory, and experiment warrants further investigations, beyond the scope of this work.

Although we experimentally report on an unequivoqual anomalous thickness-dependent conductivity scaling, and activation energy in TMH $Nb_3Cl_8$, the exact origin is still unknown. The observed trend points to an effective parallel conduction mechanism which consists of a higher conductivity surface, and lower conductivity bulk contribution. In principle, the total effective parallel conductivity can be defined as $\sigma_{eff} = \sigma_{surf} + \sigma_{bulk}$, where $\sigma_{surf}$ is the surface conductivity, and $\sigma_{bulk}$ is the bulk conductivity. The thickness independence of G implies that $\sigma_{bulk}$ is relatively low compared to $\sigma_{surf}$, which may be due to the presence of vdW interactions between adjacent layers influencing the electron-phonon scattering cross-section. $\sigma_{surf}$ is expected to be unaffected by the thickness variation, but strongly affected by imputies, and surface contamination such as oxidation or adsorbates. Since oxidation of the surface is indeed seen (but appears to be self-passivating, and only penetrating ~1 nm into the flake at most), we expect, in the very thin flakes limit, the hBN capped devices to have a smaller activation energy than the uncapped ones, which is confirmed by the 8 nm hBN-capped device exhibiting an $E_a$ of ~80 meV.

## Conclusion

In conclusion, we have measured the thickness dependent electrical conductivity of the 2D layered TMH cluster magnet $Nb_3Cl_8$. Compared to the bulk, ultra-thin flake devices (~5 nm) showed more than a three order of magnitude increase in conductivity with respect ot bulk, concomitant to a decrease in activation energy of about half of the bulk value. In addition, a monotonic decrease of the effective $E_a$ is observed as the thickness is decreased. The combination of XPS, and electrical resistivity results of hBN-encapsulated, and non-encapsulated devices allows us to conclude that surface oxidation of $Nb_3Cl_8$ did occur when suitably capped by hBN, and cannot account for the observed anomalous thickness dependent transport behavior. We hence infer that the large increase in conductivity of thin $Nb_3Cl_8$ is intrinsic, and not due to impurity or doping resulting from any of the device fabrication step. We proposed a phenomenological parallel conduction channel model consisting of a surface channel with high conductivity, and bulk channel with low conductivity to account for the observed behavior. Although the cause of the higher surface conductivity has not yet been elucidated, our observations point to a transport mechanism in this layered TMH distinct from a typical three-dimensional system. It has been theoretically predicted that the monolayer is a topological insulator (even with electron correlations included), and it may be that below 50 nm, finite size effect or remnants of higher order topology (as seen in Bi, and multilayer $WTe_2$) are responsible for the anomalous transport [36, 37, 38]. Future work studying few layers, and monolayer $Nb_3Cl_8$, particularly in the context of 2D ferromagnetism versus geometric frustration in spin liquids, as well as topology, will further our understanding of transport physics in TMH systems.

## Acknowledgements

This research was supported by the Alexander von Humboldt Foundation Sofja Kovalevskaja Award, the MINERVA ARCHES Award, the Max Plank Institute of Microstructure Physics in Halle, and the Johns Hopkins University in the USA. Jiho Yoon, Edouard Lense, and Kornelia Sklarek carried out device fabrication, measurement, and analysis. John Sheckelton, and Chris Pasco prepared the crystals. Stuart S. P. Parkin, Tyrel M. McQueen, and Mazhar N. Ali are the primary investigators.

The authors declare that they have no competing financial interests.

Correspondence, and requests for materials should be addressed to Jiho Yoon (jiho.yoon@mpi-halle.mpg.de), and Mazhar N. Ali (maz@berkeley.edu).